\documentclass[pra,twocolumn,superscriptaddress]{revtex4-1}
\pdfoutput=1

\usepackage{color}
\usepackage{graphicx}
\usepackage{dcolumn}
\usepackage{bm}
\usepackage{soul}
\usepackage{tabularx}
\usepackage{amsmath}

\usepackage{acronym}
\newacro{TVD}{Total Variation Distance}

\newcommand{\ket}[1]{\left\vert{#1}\right\rangle}

\begin{document}

\title{Experimental quantification of genuine four-photon indistinguishability}

\author{Taira Giordani}
\affiliation{Dipartimento di Fisica, Sapienza Universit\`{a} di Roma, Piazzale Aldo Moro 5, I-00185 Roma, Italy}

\author{Daniel J. Brod}
\affiliation{Instituto de F\'{i}sica, Universidade Federal Fluminense, Av. Gal. Milton Tavares de Souza s/n, Niter\'oi, RJ, 24210-340, Brazil}

\author{Chiara Esposito}
\affiliation{Dipartimento di Fisica, Sapienza Universit\`{a} di Roma, Piazzale Aldo Moro 5, I-00185 Roma, Italy}

\author{Niko Viggianiello}
\affiliation{Dipartimento di Fisica, Sapienza Universit\`{a} di Roma, Piazzale Aldo Moro 5, I-00185 Roma, Italy}

\author{Marco Romano}
\affiliation{Dipartimento di Fisica, Sapienza Universit\`{a} di Roma, Piazzale Aldo Moro 5, I-00185 Roma, Italy}

\author{Fulvio Flamini}
\affiliation{Dipartimento di Fisica, Sapienza Universit\`{a} di Roma, Piazzale Aldo Moro 5, I-00185 Roma, Italy}

\author{Gonzalo Carvacho}
\affiliation{Dipartimento di Fisica, Sapienza Universit\`{a} di Roma, Piazzale Aldo Moro 5, I-00185 Roma, Italy}

\author{Nicol\`o Spagnolo}
\email{nicolo.spagnolo@uniroma1.it}
\affiliation{Dipartimento di Fisica, Sapienza Universit\`{a} di Roma, Piazzale Aldo Moro 5, I-00185 Roma, Italy}

\author{Ernesto F. Galv\~{a}o}
\affiliation{Instituto de F\'{i}sica, Universidade Federal Fluminense, Av. Gal. Milton Tavares de Souza s/n, Niter\'oi, RJ, 24210-340, Brazil}

\author{Fabio Sciarrino}
\affiliation{Dipartimento di Fisica, Sapienza Universit\`{a} di Roma, Piazzale Aldo Moro 5, I-00185 Roma, Italy}

\begin{abstract}
Photon indistinguishability plays a fundamental role in information processing, with applications such as linear-optical quantum computation and metrology. It is then necessary to develop appropriate tools to quantify the amount of this resource in a multiparticle scenario. Here we report a four-photon experiment in a linear-optical interferometer designed to simultaneously
estimate the degree of indistinguishability between three pairs of photons. The interferometer design dispenses with the need of heralding for parametric down-conversion sources, resulting in an efficient and reliable optical scheme. We then use a recently proposed theoretical framework to quantify genuine four-photon indistinguishability, as well as to obtain bounds on three unmeasured two-photon overlaps. Our findings are in high agreement with the theory, and represent a new resource-effective technique for the characterization of multiphoton interference.
\end{abstract}

\maketitle

\section{Introduction}

Photon indistinguishability is a key concept in quantum physics, which can be characterized via two-particle interference \cite{Qinterference}. From the first experimental observation carried out using entangled photons produced by a parametric down-conversion source \cite{HOM}, the quantum signature of this phenomenon, known as Hong-Ou-Mandel (HOM) effect, has proved to be essential for tasks such as characterization of single photon sources, practical protocols of quantum communication \cite{task1,task2,task3,task4,task5,task6} and quantum computation. Recent proposals use two-photon interference for the realization of two-qubit gates, essential elements for photon-based quantum computing schemes \cite{qcphoton}. Multiphoton interference is at the core of the computational complexity of linear optical networks, as exemplified by the Boson Sampling problem \cite{AA10, Broome13boson,Spring13boson,Tillmann13boson, Crespi13boson, Spagnolo13birthday, Bentivegna15, Carolan15, Tillmann15, Wang_Pan17, Loredo17, He_Pan17, Wang18, Wang18time, Paesani18, Zhong18, Zhong19, Brod19review}. This problem, solved naturally by multiphoton interference, shows that simulating the dynamics of indistinguishable photons is likely to be hard for classical computers, which also suggests that certification of genuine multiphoton interference is a difficult problem  \cite{Aaronson14,Tichy14,Aolita15,Crespi15,Walschaers16,Bentivegna2016,Wang2016,Dittel2017,Spagnolo14,Carolan14,Bentivegna14,Crespi16,Viggianiello17optimal,Agresti17,Giordani18,Viggianiello18,FlaminiTSNE}. 

Due to the fundamental role of such quantum effects in quantum information and quantum optics, recently much effort has been devoted to increasing the number of interfering particles, and to the development of new methods to characterize the amount of indistinguishability of multiphoton states produced by single-photon sources. These methods aim to find effective strategies in terms both of physical and computational resources needed to address the problem. Many techniques are based on bosonic coalescence \cite{Spagnolo13birthday,Carolan14}, i.e. the tendency of indistinguishable photons to come out from the same port of an interferometer, which can be viewed as generalizations of the original two-photon  HOM effect to the case of multiple ports and photons \cite{TichyNJP, SpagnoloTritter, Tichy14, Carolan15,Menssen15, Crespi16, Viggianiello17optimal,ShchesnovichCircle,Agne17, Viggianiello18}.

Here we present an approach for quantifying indistinguishability in multiphoton experiments. To this aim we focus on extracting information from  sets of two-photon HOM experiments realizable in a single inteferometer with a suitable design. We describe each possible experiment by a graph, with nodes representing photons and edges representing two-photon HOM tests \cite{Brod19}. As we discuss in the following, this method provides a simple and efficient design capable of quantifying genuine multiphoton indistinguishability. Indeed, a test based on logical constraints, theoretically introduced in Ref.~\cite{Brod19overlaps}, results in lower bounds for the amount of such quantity. 
In this work we experimentally demonstrate the effectiveness of this approach, furnishing an extensive experimental study of genuine indistinguishability in a four-photon state. We also demonstrate other useful predictions of the model, which greatly simplifies the experimental effort for the complete characterization of pairwise overlaps of multiphoton states \cite{Brod19overlaps}. Simply put, the theory allows us to infer precise bounds on pairwise overlaps not accessible or measured directly by the experiment, based on information on the overlaps that were actually measured. The predictions we obtain on the unmeasured overlaps inferred from experimental data represent the first application of such techniques for a multiphoton state. By slightly changing the experimental set-up, we were able to directly confirm the theoretical prediction for one of the previously unmeasured overlaps.

Our results confirm the feasibility of the method when applied to an actual physical system. In particular, the special design of the interferometer allows to perform the experiment with a parametric down-conversion source without the need for heralding. This represents a demonstration of a practical approach to the characterization of multiphoton sources, which promises to decrease the experimental effort required to benchmark future deterministic single-photon sources  \cite{Michler2282, Ding_quantum_dot, Somaschi2016} in the regime of high number of photons \cite{Loredo17,Wang_Pan17, He_Pan17, Wang18}.

\section{Quantifying four-photon indistinguishability} \label{sec:theory}

In the recent Refs.~\cite{Brod19,Brod19overlaps}, an approach was theoretically proposed to characterize multiphoton indistinguishability, based on two-photon HOM tests and constraints imposed by logic and quantum theory. Here we review the main concepts of this approach, as applied to the case of interest for the experimental implementation we report in Section \ref{sec:exp}.

\begin{figure}[t!]
\centering
\includegraphics[width=0.99\columnwidth]{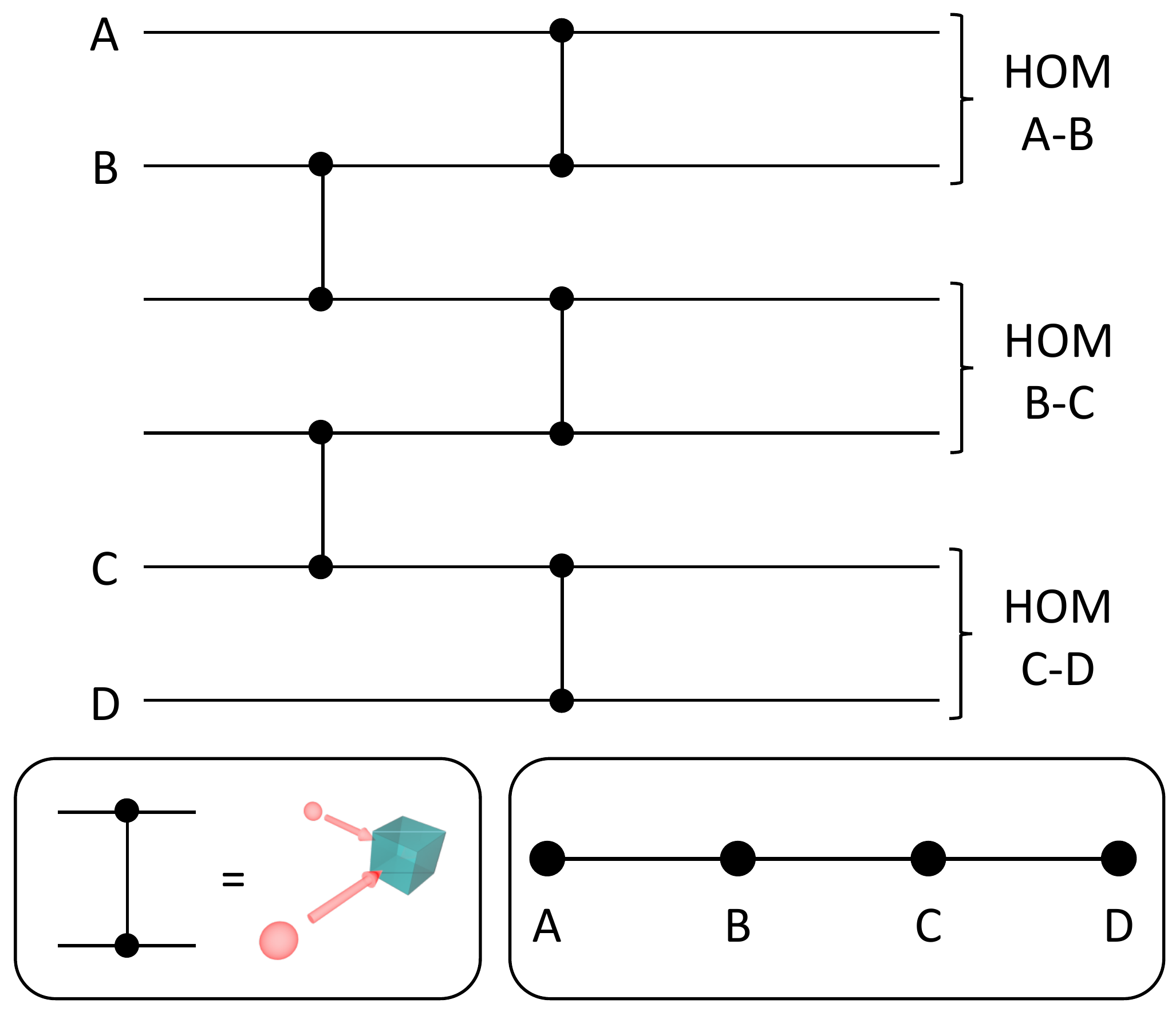}
\caption{{\bf Scheme of the four-photon indistinguishability test.} The four input photons (A,B,C,D) are injected in an interferometer which allows for the simultaneous implementation of three independent Hong-Ou-Mandel (HOM) tests between pairs (A-B), (B-C) and (C-D). In this circuit, each two-mode gate corresponds to a 50/50 beam splitter (bottom left inset). This interferometer implements pairwise HOM tests encoded in a linear graph of four nodes (bottom right inset).}
\label{fig:scheme}
\end{figure}

Consider a weighted, undirected, linear graph $P_4$ with four vertices $A, B, C$ and $D$, and three edges $r_{AB}, r_{BC}$ and $r_{CD}$ (see Fig.~\ref{fig:scheme}). Vertices correspond to single-photon states, and the edge weights to the overlap  $r_{ij}=Tr(\rho_i \rho_j)$ between the states of photons $i$ and $j$. Each overlap $r_{ij}$ can be estimated experimentally via the probability of bunching $p_{b}^{ij}$ in a HOM test between photons $i$ and $j$, as they are related by \cite{Garcia-Escartin}:
\begin{equation}
p_{b}^{ij} = \frac{1+r_{ij}}{2}.
\end{equation}
Moreover, it is possible to experimentally estimate multiple overlaps using a single interferometer, by measuring the probability of bunching at each output port. In Fig.~\ref{fig:scheme} we show such an interferometer, which we can use to estimate the three overlaps for the $P_4$ graph.

We discuss the quantification of multiphoton indistinguishability for states of two different types 
undergoing the test represented by the graph of Fig.~\ref{fig:scheme}. These two models allow us to quantify genuine indistinguishability, as well as to infer bounds on pairwise overlaps that have not been measured. The first model we consider consists of (generally mixed) four-photon states of the form: 
\begin{equation}
   \rho= c_1 \, \rho^{1} + \sum_{s > 1} c_{s} \, \rho_{s}^{\perp} , \label{eq:class}
\end{equation}
where $\rho^{1}$ is a state of 4 perfectly indistinguishable photons, and $\rho_{s}^{\perp}$ are pure states where every pair of photons is either perfectly identical or orthogonal (i.e.\ distinguishable), and at least two photons are orthogonal. In the rest of the work we will refer to this model as (\textit{i}). Following the approach of Ref.~\cite{Brod19}, we identify the degree of multiphoton indistinguishability of state (\ref{eq:class}) with the value of $c_1$, as it is the probability associated with preparing perfectly indistinguishable photons. States of the form (\ref{eq:class}) above were called ``classical'' in Ref.~\cite{Brod19overlaps}, as there exists some basis in which they are diagonal and thus display no coherences.

In Ref.~\cite{Brod19} it was shown that the HOM probabilities of bunching could be used, in a suitable family of interferometers, to obtain non-trivial bounds for $c_1$. These interferometers, which perform multiple HOM tests, can be described as star graphs, where HOM tests are represented by a central vertex connected to all the others. As shown in Supplementary Information, a similar argument can be made for the interferometer in Fig.~\ref{fig:scheme}, associated with graph $P_4$ (the possibility to find bounds for $c_1$ is implicit in the more general results of Ref.~\cite{Brod19overlaps}). More explicitly, we show that $c_1$ is bounded by:
\begin{equation}
r_{AB}+r_{BC}+r_{CD}-2 \le c_1 \le \min(r_{AB}, r_{BC}, r_{CD}). \label{eq:boundc1}
\end{equation}
Then, the measurement of $r_{AB}$, $r_{BC}$ and $r_{CD}$ via the three HOM tests in the interferometer in Fig.~\ref{fig:scheme} enables us to quantify the degree of indistinguishability.
\begin{figure*}[t!]
\centering
\includegraphics[width=0.99\textwidth]{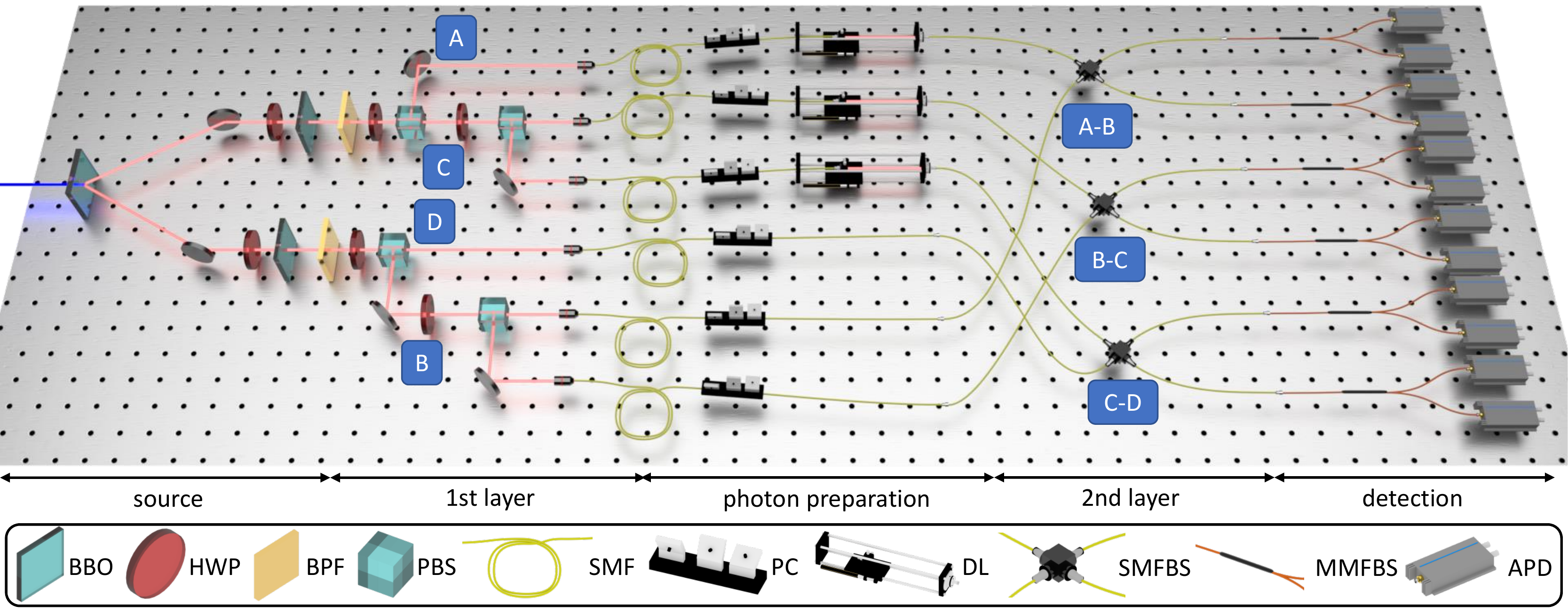}
\caption{{\bf Scheme of the apparatus used to quantify four-photon indistinguishability}. Four-photon states are generated by a type-II parametric down-conversion source in a beta-barium borate crystal, and manipulated to be indistinguishable in polarization, spectrum and temporal delay. Spatial and temporal walkoffs are compensated by means of half-length BBO crystals. With reference to Fig.~\ref{fig:scheme} photons in input modes (B,C) are split in two beam splitters, implemented here as a sequence of a half-wave plate and polarizing beam splitter for enhanced control. Photons in (B,C) separately interfere in single-mode fiber beam splitters with the other two photons in optical modes (A,D). Each output mode of the interferometer is connected to a multimode fiber beam splitter to enable approximate photon-number resolution. Legend: BBO - beta-barium borate crystal, HWP - half-wave plate, BPF - band-pass filter, PBS - polarizing beam splitter, PC - polarization controller, DL - delay lines, SMFBS - single-mode fiber beam splitter, MMFBS - multimode fiber beam splitter, APD - avalanche photodiode.}
\label{fig:expsetup}
\end{figure*}
We are interested in inferring the values of the overlaps which were not measured by our interferometer: $r_{AC}, r_{BD}$, and $r_{AD}$. In particular, it is easy to check that $c_1$ in Eq.~(\ref{eq:class}) is a lower bound for the estimate of all overlaps $r_{ij}$, hence the lower bound in Eq.~(\ref{eq:boundc1}) also applies to them. More stringent bounds can be found by applying the results of Ref.~\cite{Brod19overlaps} to the four-photon state (\ref{eq:class}) undergoing measurements at the output of the interferometer in Fig.~\ref{fig:scheme}, obtaining (see Supplementary Information):
\begin{align}
r_{AC} \in & \left[ r_{AB}+r_{BC}-1, 1-|r_{AB}-r_{BC}| \right], \\
r_{BD} \in & \left[ r_{BC}+r_{CD}-1, 1-|r_{BC}-r_{CD}|\right],\\
r_{AD} \ge & r_{AB}+r_{BC}+r_{CD}-2,\\
r_{AD} \le & 2+ \min(r_{AB}-r_{BC}-r_{CD},\\ &r_{AC}-r_{AB}-r_{CD}, r_{CD}-r_{AB}-r_{BC}) \notag.
\end{align}

We will now retrieve corresponding bounds for states described by a second model, which we label (\textit{ii}). We assume that our four-photon state  is pure and separable, i.e.\ of the form $\ket{\mathrm{A}}\otimes \ket{\mathrm{B}} \otimes \ket{\mathrm{C}} \otimes \ket{\mathrm{D}}$. We do not constrain the pure states describing different photons to be either identical or completely distinguishable. Ref.~\ \cite{Brod19overlaps} obtains general bounds for unmeasured overlaps of such product states. In the case of a four-photon state it is possible to show that:
\begin{equation}
r_{AC} \le \left( \sqrt{r_{AB}r_{BC}}+\sqrt{(1-r_{AB})(1-r_{BC})}\right)^2 \label{eq:ub}
\end{equation}
and, if $r_{AB}+r_{BC}>1$,
\begin{equation}
 r_{AC} \ge \left( \sqrt{r_{AB}r_{BC}}-\sqrt{(1-r_{AB})(1-r_{BC})}\right)^2, \label{eq:lb}
\end{equation}
otherwise $r_{AC} \ge 0$. By permuting indices $B \to A$, $C \to B$, and $D \to C$, the inequalities above also give upper and lower bounds for $r_{BD}$.
We can then prove bounds on the unknown overlap $r_{AD}$ by applying the same reasoning above, but using $r_{AB}$ and the inferred range for $r_{BD}$ (alternatively, using $r_{CD}$ and the inferred range for $r_{AC}$). These bounds are discussed in detail in Supplementary Information.

We note that the two models (\textit{i}) and (\textit{ii}) we consider are different approximations of the four-photon input state. The ideal case of four perfectly indistinguishable photons corresponds respectively to $c_1=1$ in the state described by Eq.~(\ref{eq:class}) and to state $\ket{\psi}\otimes \ket{\psi} \otimes \ket{\psi} \otimes \ket{\psi}$ in our product-state model.
The aim of this analysis is to apply these two descriptions to a four-photon state produced by a non-deterministic single-photon source, with the goal of quantifying the indistinguishability and providing estimations of unmeasured overlaps.

\section{Experimental implementation} 
\label{sec:exp}

\begin{figure*}[t!]
\centering
\hfill
\includegraphics[width=0.99\textwidth]{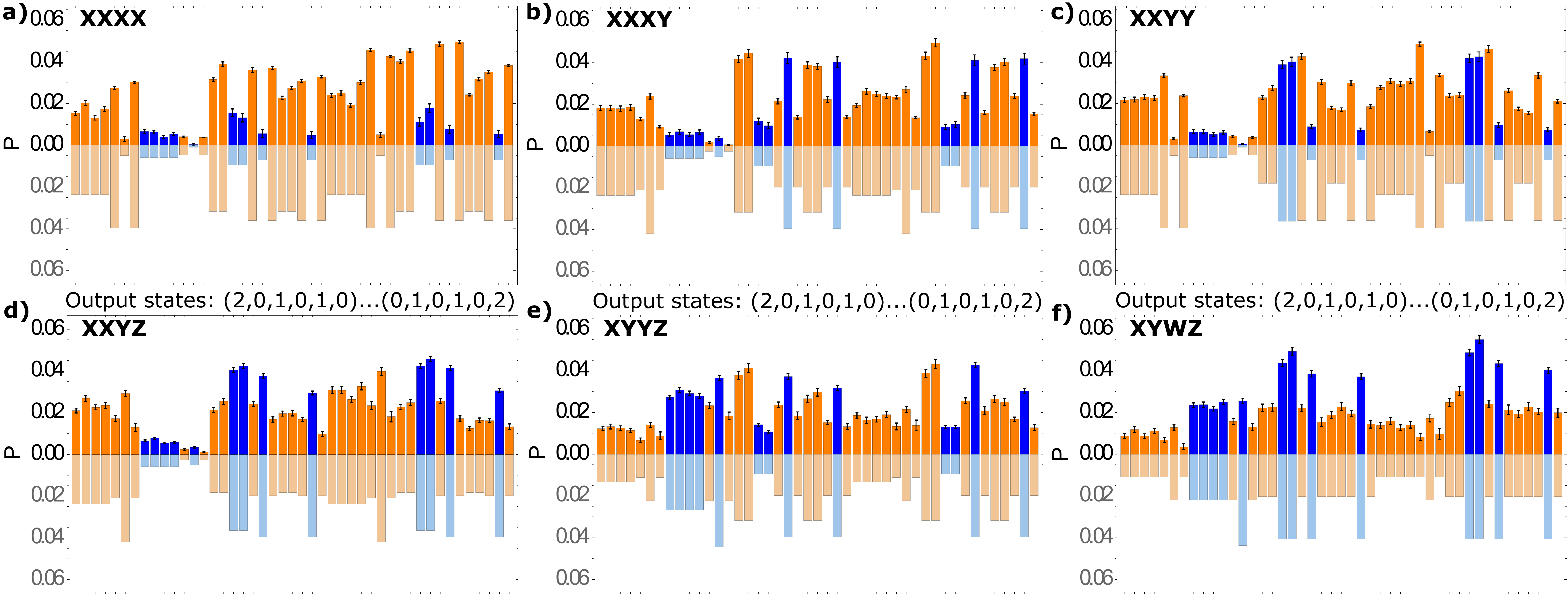}
\caption{{\bf Experimental results}. Four-photon output probability distributions for different degrees of pairwise distinguishability. Bunching and no-bunching output configurations are marked with orange and blue respectively. The experimental distributions are shown above the expected ones, reported with transparent colors. The expected distributions take into account multi-pair contributions and residual partial distinguishability among the pairs. a) Histograms for the fully indistinguishable scenario (XXXX). The measured TVD between expected and measured distribution is TVD$ =0.099(4)$. b) Configuration XXXY, corresponding to photon D distinguishable from the others (TVD$ =0.112(5)$). c) Distribution corresponding to state XXYY in which photons A-B and C-D are pairwise indistinguishable. Such condition is achieved introducing a temporal delay between photons B and C (TVD$ =0.099(3)$). d) Configuration XXYZ, where photons C-D are distinguishable (TVD $ =0.095(4)$). e) Configuration XYYZ, in which only photons B-C, belonging to different SPDC pairs, are indistinguishable (TVD $ =0.093(4)$). f) Fully distinguishable case (TVD $ = 0.110(5)$).}
\label{fig:results1}
\end{figure*}

In this section we describe the experimental apparatus we use to quantify four-photon indistinguishability (Fig.~\ref{fig:expsetup}), and also to obtain bounds for the unmeasured two-state overlaps.

To generate our input states we employ double-pair emission from a Type-II spontaneous parametric down-conversion (SPDC) source, where all four photons are spectrally filtered by means of 3 nm band-pass filters. In the first stage after photon generation, photons in input modes (B,C) propagate through a pair of beam splitters, implemented as a sequence of a half wave-plate and a polarizing beam splitter. This set of optical elements implements the first layer of the interferometer shown in Fig.~\ref{fig:scheme}. In the second stage, all 6 modes are coupled in single-mode fibers, and propagate through polarization control stages and a set of delay lines to modulate their degree of indistinguishability. In the third stage all photons,  in superposition over the six modes, are coupled to three single-mode fiber beam splitters (SMFBSs), corresponding to the second layer of the interferometer, where Hong-Ou-Mandel tests are performed. Finally, the output modes of each FBS are connected to multimode fiber beam splitters to enable approximate photon-number resolving detection. Four-photon coincidence events are then recorded via 12 single-photon avalanche photodiodes and a coincidence detection apparatus.

An important experimental characteristic of the interferometer in Fig.~\ref{fig:scheme} is that it enables four-photon experiments without the need for heralding. To understand this feature, let us start by considering only the input states corresponding to the generation of two photon pairs, which are described by the occupation numbers $\{(1,1,0,0,1,1),(2,2,0,0,0,0),(0,0,0,0,2,2)\}$. The structure of interferometer in Fig.~\ref{fig:scheme} is such that all output configurations accessible to input state $(1,1,0,0,1,1)$ are different from those associated to the other two (double pair) inputs produced by the source. This observation makes it possible to unambiguously post-select only those output events corresponding to the desired four-photon input state $(1,1,0,0,1,1)$ with no need for heralding (see Supplementary Information for further details).

The complete set of probabilities associated to all four-photon output configurations is shown in Fig.~\ref{fig:results1}. We have identified six interesting input configurations, corresponding to different situations of pairwise indistinguishability, which we labelled as \{XXXX, XXXY,  XXYY, XXYZ, XYYZ, XYWZ\} (where identical letters denote ideally indistinguishable photons). The XXXX configuration corresponds to the fully indistinguishable case, while in the other cases the number of distinguishable photons increases up to the fully distinguishable state XYWZ. We obtained fine control over pairwise distinguishability by tuning the temporal delays.

For each class of input state we collected $N \geq 10^4$ events. Fig.~\ref{fig:results1} shows the values of the distributions corresponding to the experimental data $p^\mathrm{m}$, compared against the expected distributions $p^\mathrm{e}$ that take into account partial indistinguishability and multi-pair contributions (see Supplementary Information for more details). In particular, to estimate $p^\mathrm{e}$ we considered probabilistic mixtures of states in which pairs of photons can only be either identical or perfectly distinguishable, as prescribed by model (\textit{i}). The agreement between measured $p^{\mathrm{m}}$ and expected distributions $p^{\mathrm{e}}$ is quantified using the Total Variation Distance (TVD), defined as $\mathrm{TVD} = 1/2 \sum_i \vert p^{\mathrm{m}}_i - p^{\mathrm{e}}_i \vert$ (see Fig.~\ref{fig:results1}). The comparison with the data confirms that $p^\mathrm{e}$ reproduces very well the distribution associated to the state when projected on the basis of the occupation numbers in the output modes.

\subsection{Determining genuine multiphoton indistinguishability}

In Section \ref{sec:theory} we have seen that for classical states it is possible to use measured overlaps to bound the amount of genuine multiphoton indistinguishability, as characterized by coefficient $c_1$ in Eq.~(\ref{eq:boundc1}). For the XXXX configuration, the obtained bounds for $c_1$ are $0.34 \pm 0.01\leq c_1 \leq 0.640 \pm 0.008$, indicating the presence of genuine four-photon indistinguishability by 31 standard deviations.
As expected, the other input states do not display any level of genuine four-photon indistinguishability (as $c_1=0$). Indeed, configurations different from XXXX have been measured to confirm the effectiveness of the method. The reported data show that each state with at least one photon distinguishable from the others does not display any statistically significant component of genuine multiphoton indistinguishability. The resulting bounds for $c_1$ for these states are reported in Table~\ref{table:c1}.
 
\begin{table}[h!]
   \centering
    \begin{tabular}{ccccc}
    \hline
    \textbf{State} &  $c_1$& $r_{\mathrm{AB}}$& $r_{\mathrm{BC}}$& $r_{\mathrm{CD}}$\\
    \hline
    \textbf{XXXX} &\,$[0.34(1),\,0.64(1)]$ & $0.826(6)$ & $0.640(8)$ & $0.872(4)$\\
    \textbf{XXXY} &\,$0.00(2)$ & $0.802(8)$ & $0.780(8)$ & $0.00(2)$\\
    \textbf{XXYY} &\,$0.01(1)$ & $0.832(6)$ & $0.01(1)$ & $0.802(6)$\\
    \textbf{XXYZ} &\,$0.00(1)$ & $0.834(6)$ & $0.00(1)$ & $0.01(1)$\\
    \textbf{XYYZ} &\,$0.01(2)$ & $0.01(2)$ & $0.68(1)$ & $0.04(2)$\\
    \textbf{XYWZ} &\,$0.00(2)$ & $0.02(2)$ & $0.00(2)$ & $0.00(2)$\\
    \hline
    \end{tabular}
     \caption{\textbf{Summary of the measured overlaps and inferred bounds for $\bm{c_1}$.} Only state XXXX displays a significant non-zero value for $c_1$. For the other states, in which at least one photon is distinguishable from the others, the inferred value for $c_1$ is statistically compatible with $c_{1}=0$. For these states we have reported the value of the upper bound in Eq.~(\ref{eq:boundc1}). The overlap values have been retrieved from the distributions shown in Fig.~\ref{fig:results1}. All the errors are derived from Poissonian uncertainties associated to photon counts and then propagated via Monte Carlo methods.}
     \label{table:c1}
\end{table}

\begin{figure}[t]
\centering
    \includegraphics[width=0.49\textwidth]{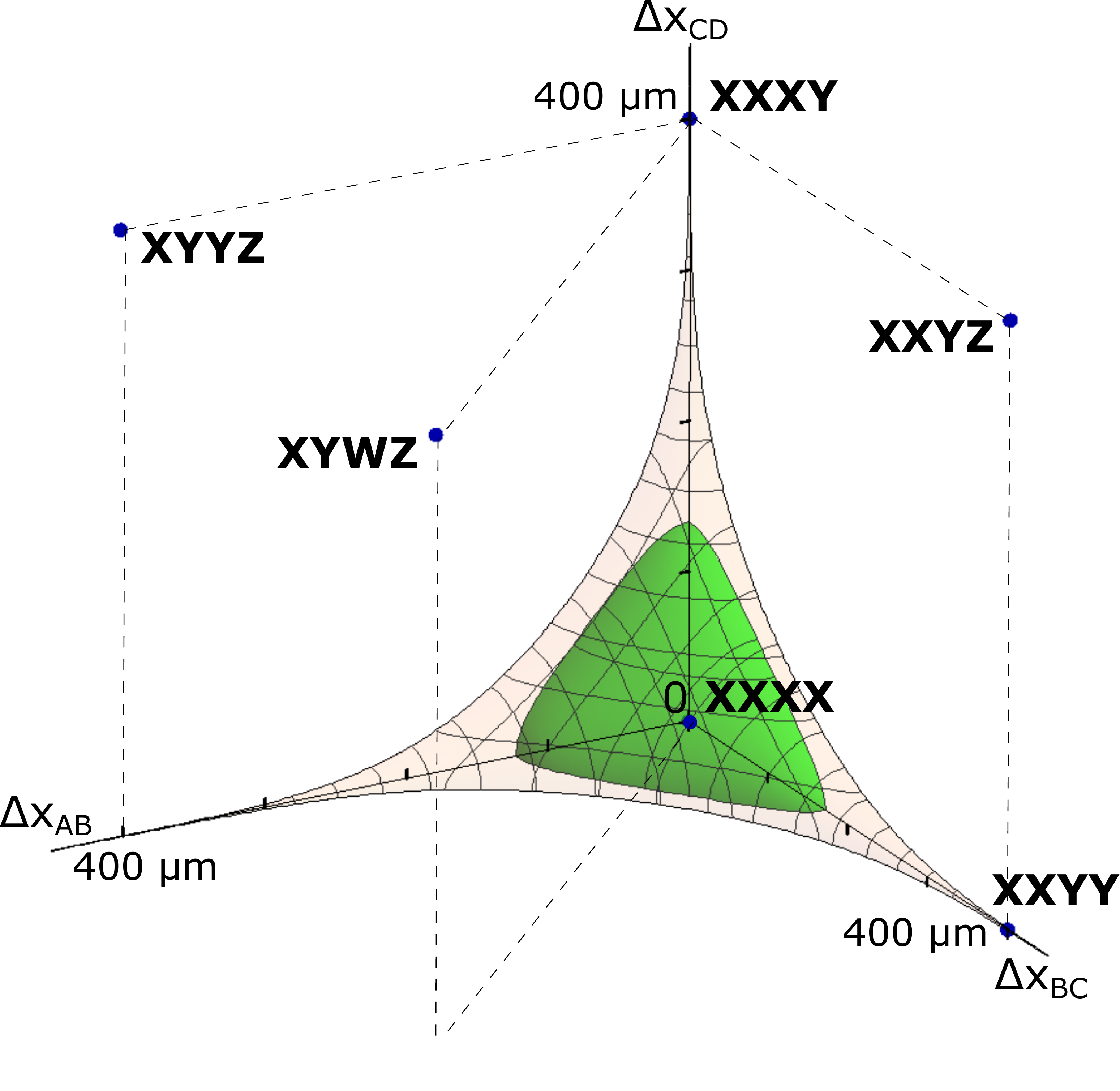}%
 \caption{\textbf{Surfaces of trivial lower bound for $\bm{c_1}$.} Input states $\{$XXXX, XXXY, XXYY, XXYZ, XYYZ, XYWZ$\}$ are represented in the space of relative optical delays between photon pairs $(i,j)=\{$(A,B), (B,C), (C,D)$\}$. The pink and green surfaces represent parameter values yielding a trivial lower bound for $c_1$, i.e $c_1=0$. The pink surface has been calculated for the case of Gaussian shape for the overlaps $r_{ij}(\Delta x_{ij})$ (inferred from the HOM dips), considering perfect indistinguishability, $r_{ij}(0)=1$. The green surface takes into account the actual observed degree of distinguishability in the experiment. Here we have considered $r_{ij}(0)=V_{ij}$, where $V_{ij}$ is the visibility of the corresponding HOM dip.}
 \label{fig:surface}
\end{figure}

In Fig.~\ref{fig:surface}, each configuration has been represented in the space spanned by the three relative delays $(\Delta x_{\mathrm{AB}}$, $\Delta x_{\mathrm{BC}}$, $\Delta x_{\mathrm{CD}})$. The plot shows the predicted surfaces corresponding to the lower bound for $c_1 $, defined in Eq.~(\ref{eq:boundc1}), calculated for Gaussian overlaps ($r_{\mathrm{AB}}$, $r_{\mathrm{BC}}$, $r_{\mathrm{CD}}$) with respect to the relative delays in the ideal case (pink surface) and by considering partial photon indistinguishability corresponding to the adopted source (green surface). For each of the two scenarios, all points enclosed within the corresponding surface lead to a non-trivial lower bound for $c_1$, which characterizes genuine multiphoton indistinguishability. The parameters adopted for the calculation of the two surfaces are inferred from the widths and visibilities of the HOM dips (further details on HOM measurements and on the estimation of the surfaces are given in Supplementary Information). The ratio between the volumes enclosed in the two surfaces, $V_a$ for the pink one and $V_b$ for the green, is $V_b/V_a\sim 0.525$, thus confirming that the region corresponding to a non-trivial lower bound for $c_1$ is smaller when experimental imperfections are taken into account.

\subsection{Determining unmeasured overlaps}

Our experimental apparatus measures a set of three two-photon overlaps: $r_{AB}, r_{BC}$ and $r_{CD}$. The theoretical results described in Section \ref{sec:theory} allow us to obtain bounds on the unmeasured overlaps $r_{AC}, r_{BD}$ and $r_{AD}$. As discussed previously, this can be done assuming two different models, (\textit{i}) and (\textit{ii}), for the states we prepare. The first model assumes that the state is classical, i.e. a  coherence-free mixture of states for which every pair of photons is either distinguishable or identical to each other. The second model assumes that the four photons are in a tensor product of pure states with any degree of two-photon indistinguishability. It is worth noting that both models could reproduce well the behaviour of the generated four-photon states in the basis of mode occupation numbers, and can thus be employed to infer predictions on the four-photon state indistinguishability. In Table~\ref{tab:unm_overlaps} we report the calculated upper and lower bounds for unmeasured overlaps, using the two models. As we can see, for all cases where unmeasured overlaps were expected to be non-zero, we obtained lower bounds consistent with this expectation. This includes positive values for the three unmeasured two-photon overlaps for input XXXX, as well as a positive value for $r_{AC}$ for input XXXY.

\begin{figure}
    \centering
    \includegraphics[width=0.46\textwidth]{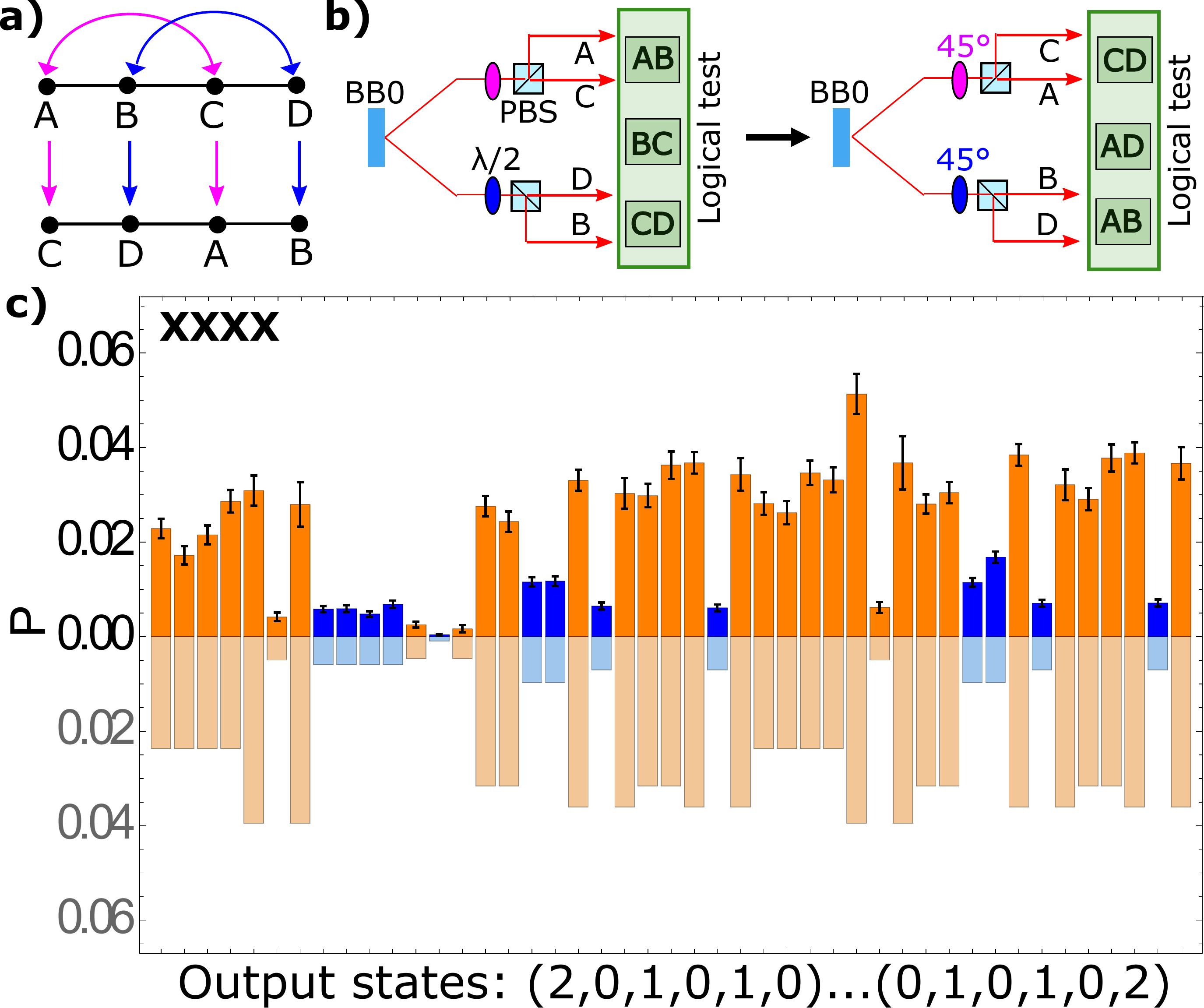}
    \caption{\textbf{Graph C-D-A-B}. a) By swapping photon pairs A/C and B/D, we do measurements corresponding to the linear graph C-D-A-B. b) In the experimental set-up, the swap is performed by rotating the half-waveplates in each arm of the source by 45$^{\circ}$ with respect to the set-up corresponding to the A-B-C-D graph. In this way, photons A,B,C,D are injected in different input ports of the same interferometer. c) Output distribution for the XXXX state of graph C-D-A-B for $N \sim 5200$ events collected. The TVD with respect to the expected distribution, reported below the experimental data with transparent colors, is TVD=$0.078(6)$.}
    \label{fig:CDAB}
\end{figure}

\begingroup
\setlength{\tabcolsep}{2pt} %
\begin{table*}[ht!]
    \centering
    \begin{tabular}{cccc|ccc}
    \hline
    \textbf{State}   &\multicolumn{3}{c|}{Mixed classical state model (\textit{i})}& \multicolumn{3}{c}{Separable pure state model (\textit{ii})} \\
            & $r_{\mathrm{AC}}$ & $r_{\mathrm{AD}}$ & $r_{\mathrm{BD}}$  & $r_{\mathrm{AC}}$ & $r_{\mathrm{AD}}$& $r_{\mathrm{BD}}$ \\
    \hline
    \textbf{XXXX}  & $\big [0.46(1),\, 0.81(1)\big ]$ & $\big [0.34(1), \, 0.94(1)\big ]$ & $\big [0.511(9), \, 0.768(9)\big ]$  & $\big [0.23(1), \, 0.955(5)\big ]$ &  $\big [0.017(4),\,0.976(4)\big ]$ & $\big [0.28(1), \, 0.925(6)\big ]$\\
    \textbf{XXXY}  & $\big [0.58(1),\, 0.98(1)\big ]$ & $\big [0, \, 0.40(2)\big ]$ & $\big [0, \, 0.20(2)\big ]$  & $\big [0.34(1), \, 0.9991(8)\big ]$ &  $\big [0,\,1\big ]$ & $\big [0, \, 1\big ]$\\
    \textbf{XXYY}  & $\big [0, \, 0.17(1)\big ]$ & $\big [0, \, 0.37(1)\big ]$ & $\big [0, \, 0.20(1)\big ]$  & $\big [0, \, 1\big ]$ &  $\big [0,\,1\big ]$ & $\big [0, \, 1\big ]$\\
    \textbf{XXYZ}  & $\big [0, \, 0.14(1)\big ]$ & $\big [0, \, 1\big ]$ & $\big [0, \, 0.97(2)\big ]$  & $\big [0, \, 1\big ]$ &  $\big [0,\,1\big ]$ & $\big [0, \, 1\big ]$\\
    \textbf{XYYZ}  & $\big [0, \, 0.33(1)\big ]$ & $\big [0, \, 1\big ]$ & $\big [0, \, 0.36(2)\big ]$  & $\big [0, \, 1\big ]$ &  $\big [0,\,1\big ]$ & $\big [0, \, 1\big ]$\\
    \textbf{XYWZ}  & $\big [0, \, 0.94(3)\big ]$ & $\big [0, \, 1\big ]$ & $\big [0, \, 0.96(2)\big ]$  & $\big [0, \, 1\big ]$ &  $\big [0,\,1\big ]$ & $\big [0, \, 1\big ]$\\
    \hline
    \end{tabular}
    \caption{\textbf{Bounding unmeasured overlaps.} We report the lower bounds (LBs) and upper bounds (UBs) for the unmeasured overlaps starting from those observed in the experiment for each state. We adopt two different models for describing the multiphoton state, the first expressed in Eq.~(\ref{eq:class}) and the second in the form $\ket{A}\otimes \ket{B}\otimes \ket{C}\otimes \ket{D}$. The bounds reported without uncertainty correspond to the cases for which the model gives trivial bounds, i.e. 0 for LBs and 1 for UBs. 
    }
    \label{tab:unm_overlaps}
\end{table*}
\endgroup
The interferometer we have implemented allows, with a slight change in the set-up, to verify directly the prediction of Table~\ref{tab:unm_overlaps} regarding the overlap $r_{AD}$. This is possible by swapping photons A with C and B with D at the input of the interferometer, by acting on the polarization degree of freedom of the source (see Fig.~\ref{fig:CDAB}). This operation results in measurements corresponding to the graph C-D-A-B, all performed with input configuration XXXX. The new measurements for overlaps  $r_{AB}, r_{CD}$ resulted in values  $r_{\mathrm{CD}}=0.86(1)$, $r_{\mathrm{AB}}=0.842(8)$. Interestingly, we can now estimate also the previously unmeasured $r_{AD}$ [$r_{\mathrm{AD}}=0.65(1)$]. The new measurements of overlaps $r_{AB}$ and $r_{CD}$ are compatible with the values obtained with the previous set-up, while the new measured value for $r_{AD}$ is compatible with the previously derived upper and lower bounds.

\section{Discussion} 

Multiphoton indistinguishability is a promising resource for quantum information processing, and so it is crucial to identify genuine instances of it, as well as characterize it efficiently \cite{Brod19, Brod19overlaps}. Here we have experimentally demonstrated an approach to quantify genuine multiphoton indistinguishability in a four-photon experiment. We also characterize the degree of two-photon indistinguishability without the need for direct measurements of all two-photon experiments.
A fundamental advantage of our approach is the use of a single interferometric set-up to effectively implement multiple Hong-Ou-Mandel tests. The experimental data, besides this direct characterization of the methodology introduced in Refs.~\cite{Brod19, Brod19overlaps}, can be used to bound all the unmeasured two-photon overlaps. Our interferometer design allows for the use of multiple parametric down-conversion single-photon sources, without the need for heralding. These features showcase the promise of our approach for the characterization of future multiphoton sources. In particular, the method could find applications in the case of deterministic sources \cite{Michler2282, Ding_quantum_dot, Somaschi2016, Huber18, Basso19}, which have been recently identified as a promising route to obtain larger photon numbers with high purity \cite{Loredo17,Wang_Pan17, He_Pan17, Wang18}.

\begin{acknowledgments}
We acknowledge funding from the European Research Council (ERC) Advanced Grant CAPABLE (Composite integrated photonic platform by femtosecond laser micro-machining, Grant Agreement No. 742745), from Sapienza Progetto di Ateneo H2020-ERC QUMAC (MAChine learning via hybrid integrated QUANTUM photonics), and from Instituto Nacional de Ci\^encia e Tecnologia de Informa\c{c}\~ao Qu\^antica (INCT-IQ/CNPq). G.C. acknowledges Becas Chile and Conicyt.
\end{acknowledgments}

\end{document}


\title{Experimental quantification of genuine four-photon indistinguishability: supplementary information}

\author{Taira Giordani}
\affiliation{Dipartimento di Fisica, Sapienza Universit\`{a} di Roma, Piazzale Aldo Moro 5, I-00185 Roma, Italy}

\author{Daniel J. Brod}
\affiliation{Instituto de F\'{i}sica, Universidade Federal Fluminense, Av. Gal. Milton Tavares de Souza s/n, Niter\'oi, RJ, 24210-340, Brazil}

\author{Chiara Esposito}
\affiliation{Dipartimento di Fisica, Sapienza Universit\`{a} di Roma, Piazzale Aldo Moro 5, I-00185 Roma, Italy}

\author{Niko Viggianiello}
\affiliation{Dipartimento di Fisica, Sapienza Universit\`{a} di Roma, Piazzale Aldo Moro 5, I-00185 Roma, Italy}

\author{Marco Romano}
\affiliation{Dipartimento di Fisica, Sapienza Universit\`{a} di Roma, Piazzale Aldo Moro 5, I-00185 Roma, Italy}

\author{Fulvio Flamini}
\affiliation{Dipartimento di Fisica, Sapienza Universit\`{a} di Roma, Piazzale Aldo Moro 5, I-00185 Roma, Italy}

\author{Gonzalo Carvacho}
\affiliation{Dipartimento di Fisica, Sapienza Universit\`{a} di Roma, Piazzale Aldo Moro 5, I-00185 Roma, Italy}

\author{Nicol\`o Spagnolo}
\email{nicolo.spagnolo@uniroma1.it}
\affiliation{Dipartimento di Fisica, Sapienza Universit\`{a} di Roma, Piazzale Aldo Moro 5, I-00185 Roma, Italy}

\author{Ernesto F. Galv\~{a}o}
\affiliation{Instituto de F\'{i}sica, Universidade Federal Fluminense, Av. Gal. Milton Tavares de Souza s/n, Niter\'oi, RJ, 24210-340, Brazil}

\author{Fabio Sciarrino}
\affiliation{Dipartimento di Fisica, Sapienza Universit\`{a} di Roma, Piazzale Aldo Moro 5, I-00185 Roma, Italy}

\maketitle

This supplementary material provides some further details both on the experimental and the theoretical background of the work. It is structured as follows. The first section illustrates the characteristics of the experimental platform, with comprehensive details about the single-photon source, the interferometer and the detection. Then we describe the preliminary characterization of the state generated by the source. The third section analyzes the post-selection of the output states. The fourth and fifth sections describe the model employed to calculate the expected distribution, taking into account the noise due to multi-pair emission from the source and the partial photon indistinguishability. In the last section we describe how we can estimate bounds on the genuine indistinguishability coefficient $c_1$, as well as how to obtain lower and upper bounds for the unmeasured overlaps.

\section{Experimental implementation}
\label{sec:exp_impl}
In this section we describe the experimental apparatus used to quantify four-photon indistinguishability, shown in Fig. 2 of the main text.

\begin{figure*}[ht!]
\centering
\includegraphics[width=0.99\textwidth]{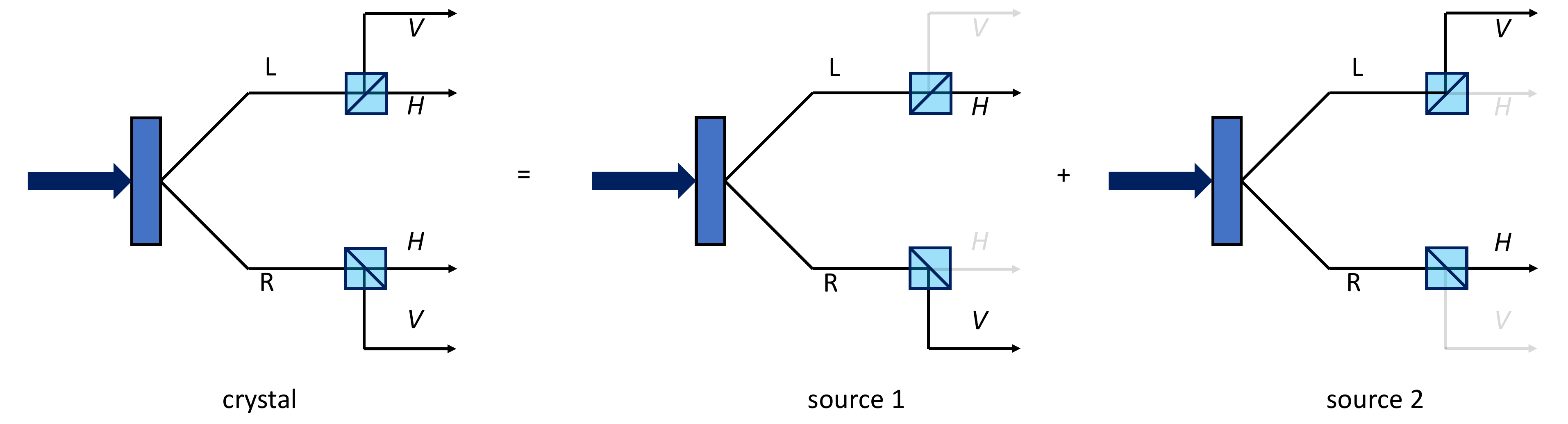}
\caption{{\bf Scheme of the employed photon generation apparatus.} Two photon-pair sources (source 1 and source 2) are embedded in a single non-linear crystal, and are separated in different spatial modes by using polarizing beam splitter.}
\label{fig:source}
\end{figure*}

\begin{figure*}[ht!]
\centering
\includegraphics[width=0.99\textwidth]{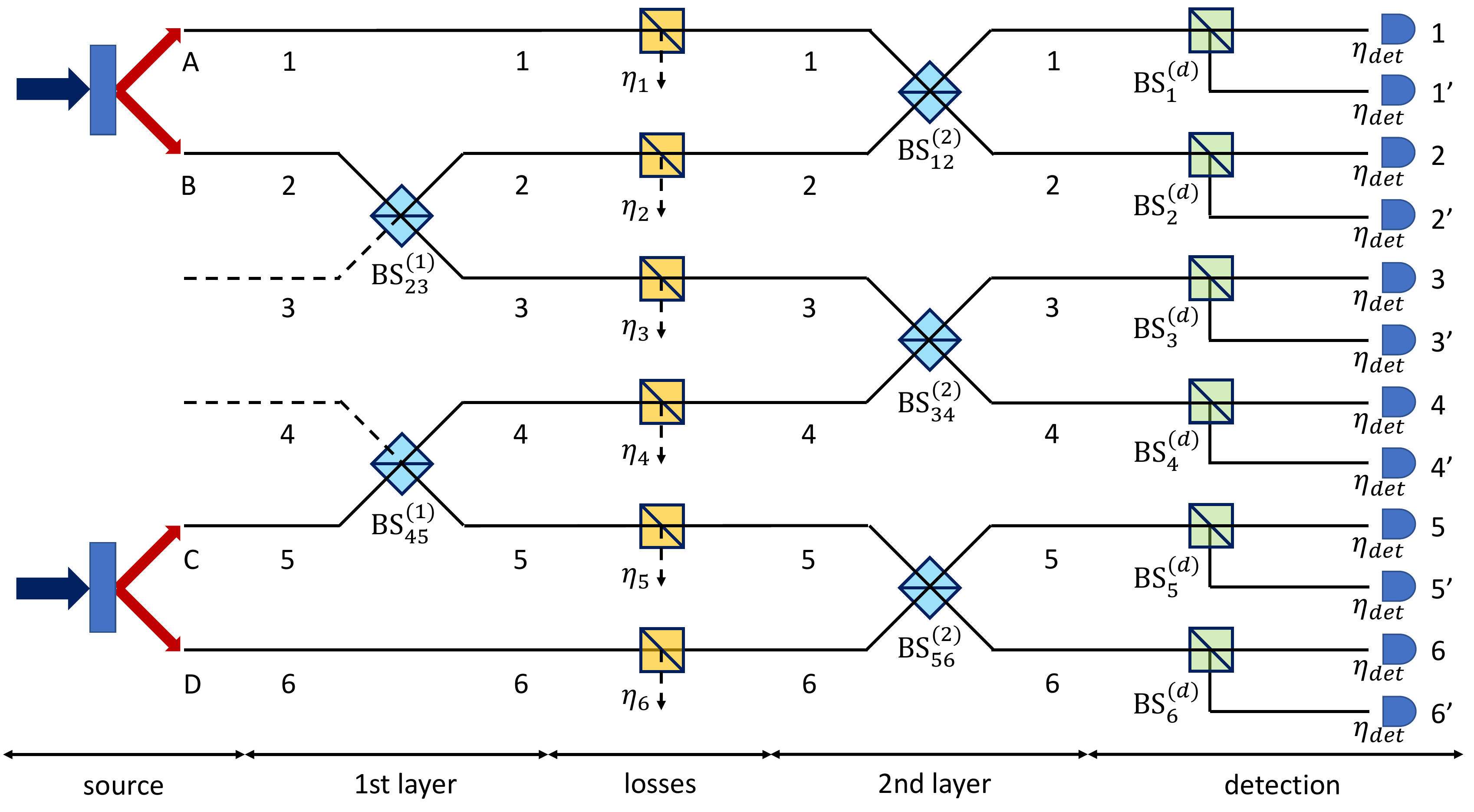}
\caption{{\bf Scheme of the adopted model for the experimental results.} The main stages are described in Sec. \ref{sec:model} and comprise photon generation, beam splitters layer 1, losses, beam splitters layer 2, and detection.}
\label{fig:model}
\end{figure*}

Let us first discuss the apparatus adopted for single-photon generation. Single photons were generated by two parametric down conversion (PDC) sources, occurring in a single nonlinear crystals (BBO) injected by a 600 mW pulsed pump field ($\lambda= 392.5$ nm). The generated two-photon pairs centered at 785 nm were filtered by 3 nm interferential filters (Semrock). Each photon source generates photons with orthogonal polarizations [horizontal ($H$) and vertical ($V$) polarizations] in two spatial modes $L$ and $R$, according to the configuration shown in Fig. \ref{fig:source}. Photons from the different sources are then separated in different spatial modes by means of polarizing beam splitter (PBS). The output state of each source is:
\begin{equation}
\ket{\psi_{\text{out}}} = \frac{1}{\cosh g} \sum_{n} \tanh^{n}(g) \vert n,n \rangle
\end{equation} 
where $n$ is the number of the generated pairs. The quantity $g$ is smaller then $1$ ($g \sim 0.1$ in our case), such that the creation of a large number of photon pairs is negligible. The state employed in our test corresponds to a four-photon generation event, when each source generates a single photon pair. We use the post-selection to select the simultaneous generations of a photon pair in each source, and to separate those events corresponding to different input states (see Sec. \ref{sec:post_sel}). By using this configuration, one source injects photons A and B in modes 1 and 2 of the interferometer, while the other source injects photons C and D in modes 5 and 6 (see Fig. \ref{fig:model} for a scheme of the apparatus equivalent to the actual set-up shown in Fig. 2 of the main text).

The first layer of the interferometer is realized by inserting two beam splitters (BS) on input modes 2 and 5. In this way the four-photon input is probabilistically split to six modes (1,2,3,4,5,6). Each BS of the first layer is realized with a half-wave plate and a PBS. In this way we obtain a BS with adjustable reflectivity, leading to the capability of setting the transmittivity to $50 \%$ for each BS. Then, all 6 modes are coupled to single-mode fibers.

The second interferometer layer is obtained by connecting mode pairs 1-2, 3-4, and 5-6 to three single-mode fiber beam splitters (SMFBS), where Hong-Ou-Mandel tests are performed. Before this stage, it is necessary to control indistinguishability of the photons in all degrees of freedom. In order to control the input polarization of each incoming photon, we employed a polarization control stage on each mode. Time differences between input photon paths are adjusted via delays inserted on one input mode of each SMFBS. The output modes of each SMFBS correspond to the six output modes of the interferometer.

Finally, detection is performed by avalanche photodiodes (APD), employed to measure the output state by counting the coincidences between the detectors. Before the detectors, each output $j$ is connected to a multimode fiber beam splitter (MMFBS), used to add photon number resolution capabilities to the apparatus. Indeed, when two photons are injected in the MMFBS on mode $j$, the latter separates probabilistically the two photons on the output modes $j$ and $j'$. A coincidence between modes $j$ and $j'$ corresponds to the detection of two photons on output mode $j$ of the interferometer.

One advantage of the described interferometer is the possibility of performing the four-photon tests without the need for a heralding photon. Indeed, in order to select the desired initial state, we can apply post-selection to the experimental data. This procedure is described in Sec. \ref{sec:post_sel}.

\section{Preliminary characterization of photon indistinguishability}
\label{sec:part_ind}

In this section we report the preliminary characterization of the actual degree of indistinguishability in the four-photon state produced by the source. 

\begin{figure*}
\centering
\includegraphics[width=\textwidth]{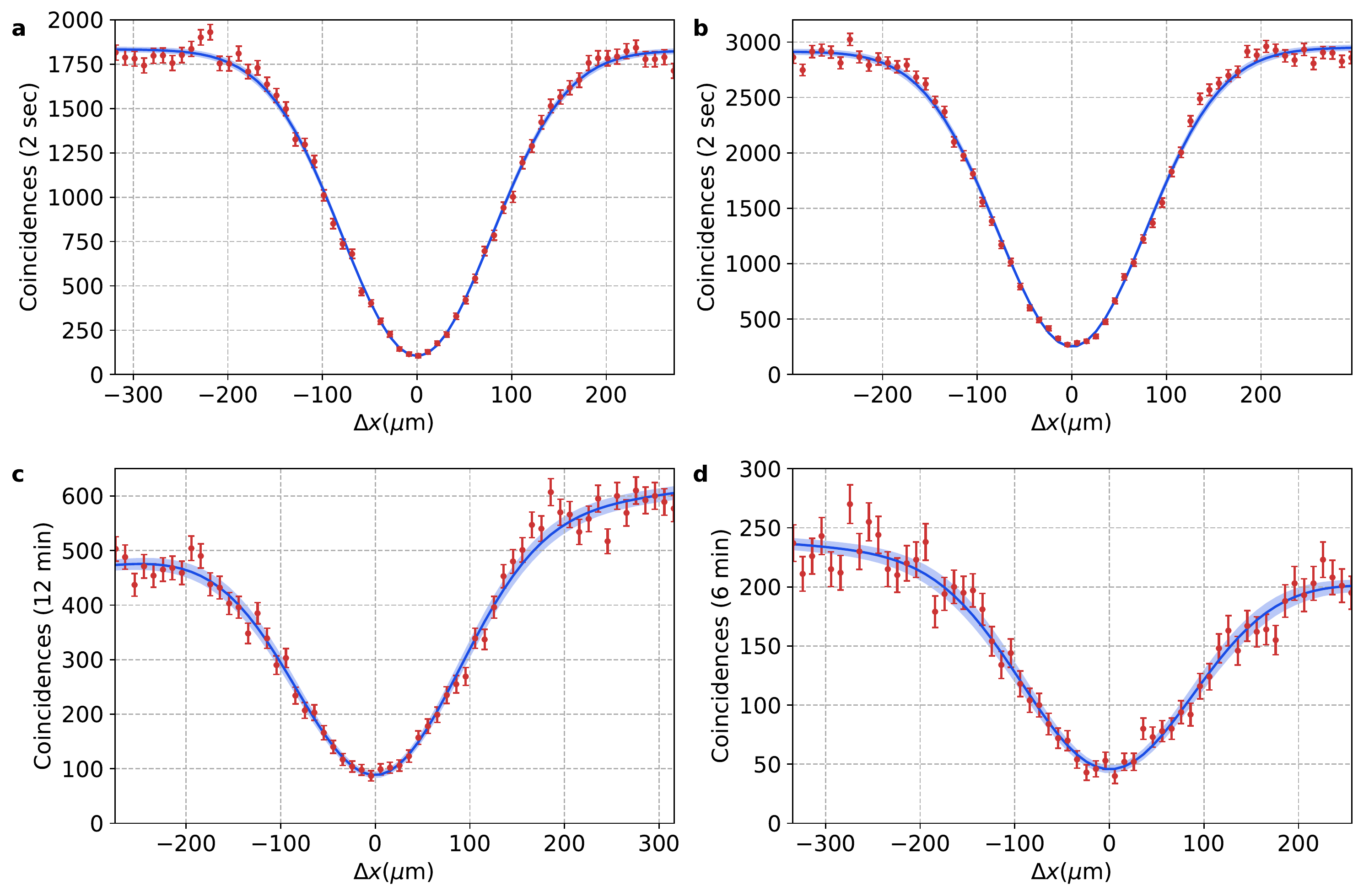}
\caption{{\bf HOM interference patterns measured in a 50/50 beam splitter for different photon combinations.} {\bf a-b}, HOM dip for photons (A,B) ({\bf a}) and (C,D) ({\bf b}) belonging to the same photon pair, as a function of the relative delay $\Delta x$. Data are obtained by measuring two-fold coincidences at the beam splitters output. The measured dip visibilities are respectively $V_{\mathrm{AB}} = 0.944(3)$ and $V_{\mathrm{CD}} = 0.915(3)$. {\bf c-d}, HOM dip photons (B,C) ({\bf c}) and (A,D) ({\bf d}) belonging to different photon pairs, as a function of the relative delay $\Delta x$. Data are obtained by measuring four-fold coincidences between the beam splitters output and the other two generated photons acting as a trigger. The measured dip visibilities are respectively $V_{\mathrm{BC}} = 0.835(7)$ and $V_{\mathrm{AD}} = 0.790(11)$. In all plots, error bars are due to the Poissonian statistics of the measured coincidences. Solid lines are best fit of the experimental data according to the function $C(\Delta x) = A (1-B \Delta x) \{1-V \exp[-(\Delta x - \Delta x_{0})^{2}/\sigma^2]\}$, where $V$ is the visibility and the term $(1-B \Delta x)$ takes into account a decrease (increase) in the input power during the measurement. Shaded regions correspond to 1 standard deviation on the fit parameters.} 
\label{fig:dips}
\end{figure*}

The indistinguishability characterization proposed in this work is based on measuring bunching probabilities $p_b$ in Hong-Ou-Mandel (HOM) interference experiments. The bunching probability $p_b$ is related to the overlap $r$ between two photons ($p_b=\frac{1+r}{2}$), which in turn corresponds to the visibility $V$ of HOM dips. To obtain a preliminary characterization of the source we have performed independent HOM interference experiment for pairs $\{$(A,B), (B,C),(C,D), (A,D)$\}$, by injecting the different photon pair combinations directly in the input port of a balanced fiber beam splitter. The measured interference fringes, are reported in Fig.\ref{fig:dips}. The pairs $\{$(A,B), (C,D)$\}$ are produced by the same pump-photon, so that the HOM dips can be retrieved by recording two-fold coincidences. Pairs $\{$(B,C), (A,D)$\}$, correspond to one photon from each source, thus requiring four-fold coincidence detection. The resulting visibilities suggest that the four-photon state has a residual degree of distinguishability deriving from spatial and spectral correlations present in photons belonging to the same pair. In particular, such correlations are reflected in the lower visibility ($V_{\mathrm{BC},\mathrm{AD}}< 0.90$) of HOM dips between photons belonging to different pairs.

\section{Identification of the output distribution}
\label{sec:post_sel}
\begin{figure*}
\centering
\includegraphics[width=0.99\textwidth]{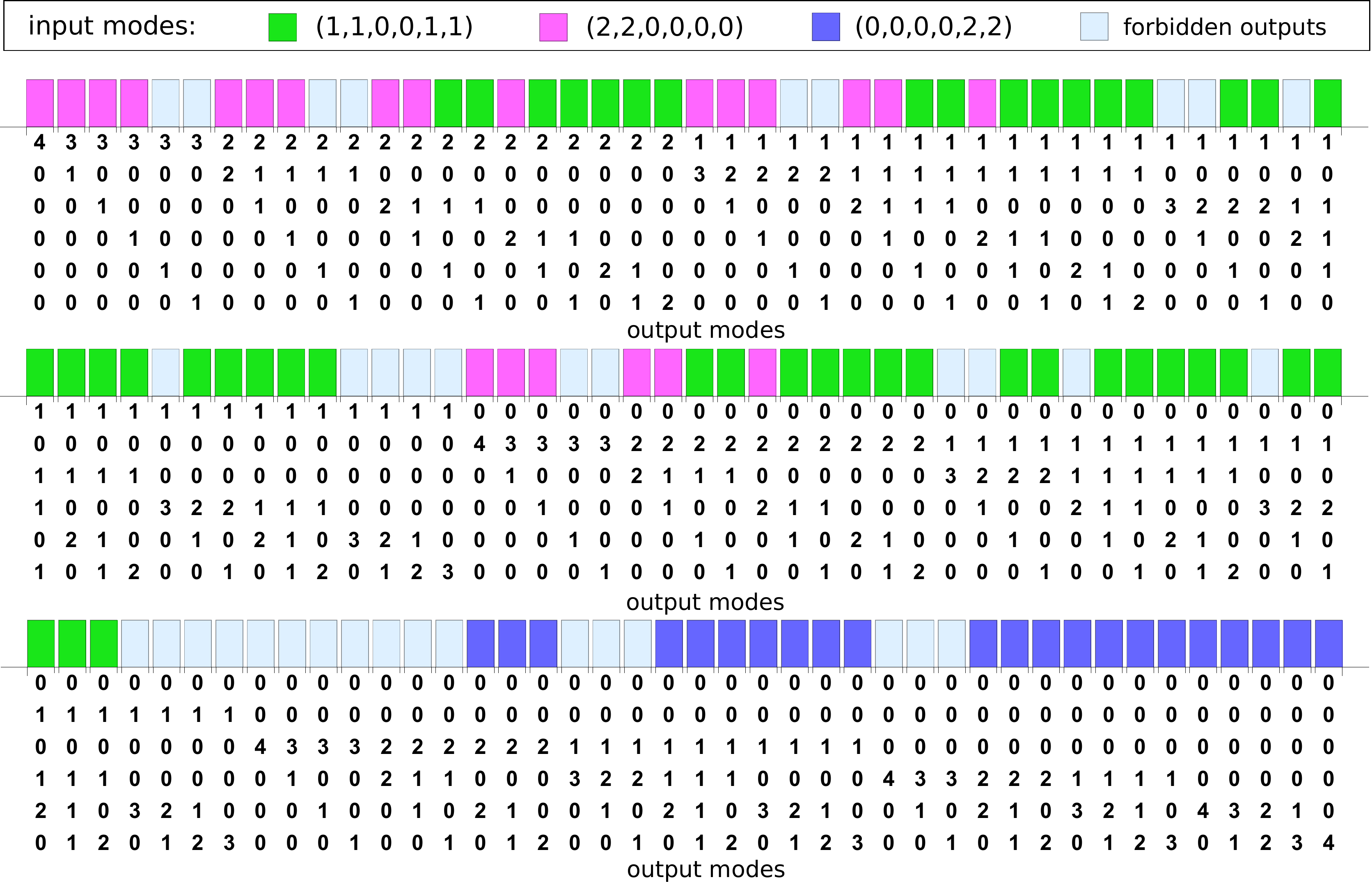}
\caption{{\bf Separability of the interferometer output configurations with respect to the four-photon input states.} Diagram showing the possible output configurations of the six-mode interferometer, shown on the x-axis. Each output state can be generated only by a different input. In the diagram, colors map each output to the corresponding input that can generate it [green: (1,1,0,0,1,1), purple: (2,2,0,0,0,0), blue: (0,0,0,0,2,2)]. Forbidden output configurations (cyan) cannot be reached with the input states generated by the experimental apparatus.} 
\label{fig:hilbertspace}
\end{figure*}

In this section we describe the post-selection of the experimental data to identify those events corresponding to the correct input state. Such post-selection procedure allows us to verify which measured events are related to an input state with one photon for each mode of the interferometer, without the need for heralding. Given the photon generation apparatus described previously and in the main text, three four-photon input states are possible. These three input states correspond to the following configurations (see Fig. \ref{fig:source}) $H_{R}H_{R}V_{L}V_{L}$, $V_{R}V_{R}H_{L}H_{L}$ or $H_{R}V_{R}V_{L}H_{L}$. If we denote the input states as $(n_1,n_2,n_3,n_4,n_5,n_6)$, where $n_i$ is the number of photons in mode $i$, the three possible configurations correspond to the input states $(2,2,0,0,0,0)$, $(0,0,0,0,2,2)$ and $(1,1,0,0,1,1)$ respectively. Post-selection of the experimental data is performed by observing that the output configurations generated by the these input states are disjoint, as shown in Fig. \ref{fig:hilbertspace}. By exploiting this property, it is possible to uniquely determine the input state if a given output configuration is observed. In other words, there is a one-to-one correspondence between possible inputs and observed output configurations. In the experiment, we selected those outputs corresponding to the requested input state (1,1,0,0,1,1), that are marked in green in Fig.~\ref{fig:hilbertspace}.  

\section{Model of the experimental distributions}
\label{sec:model}

Here we discuss a theoretical model describing the measured distributions. The corresponding experimental apparatus is shown in Fig. 2 of the main text. The main sources of imperfections in the experimental apparatus are partial photon indistinguishability, higher order photon number emission terms, non-ideal BS transmittivities, losses and limited detection efficiency. A full scheme of the adopted modelization is shown in Fig. \ref{fig:model}. 

The photon source, as discussed in Sec. \ref{sec:exp_impl}, corresponds to two independent sources emitting photon pairs on input ports (1,2) and (5,6). The main term is obtained from the emission of one photon pair per source, thus corresponding to one photon for each of the four injected input form: (1,1,0,0,1,1). As discussed in Sec. \ref{sec:part_ind}, the emitted photons present a non-unitary degree of indistinguishability, quantified by the visibility of the HOM interference pattern directly measured after a 50/50 beam splitter. Hence, the evolution for each pair of photons $(i,j)$ interfering in layer 2 of the interferometer (see Fig. \ref{fig:model}) can be modelled by a two-photon density matrix of the form $\rho_{ij} = V_{ij} \rho_{\mathrm{ind}} + (1-V_{ij}) \rho_{\mathrm{dis}}$. Here, $\rho_{\mathrm{ind}}$ stands for a state with two fully indistinguishable photons, $\rho_{\mathrm{dis}}$ stands for a state with two fully distinguishable particles, and $V_{ij}$ is the two-photon visibility.

Given the probabilistic nature of the photon source, there is also a non-zero probability that higher order emission terms are generated. More specifically, while the four-photon state (1,1,0,0,1,1) is generated with probability $p_{(1,1,0,0,1,1)}= \tanh(g)^4/\cosh(g)^4$, six photon contributions (2,2,0,0,1,1) and (1,1,0,0,2,2) are generated with probability $p_{(2,2,0,0,1,1)} = p_{(1,1,0,0,2,2)} = \tanh^6(g)/\cosh^4(g)$. In our source, the parameter $g$ has been estimated from the number of generated photons and detected coincidences, having a value $g \sim 0.1$. Given the presence of losses and non-photon number resolving detectors, those six-photon terms can contribute to the output distribution from which the bunching probabilities are estimated.  

Other sources of imperfections are given by the interferometric structure. Layer 1 of the interferometer has been implemented by half wave plates and PBSs, as shown in Fig. 2 of the main text. Hence, the reflectivies of beam splitter BS$_{23}^{(1)}$ and BS$_{45}^{(1)}$, can be considered to be $R\sim 1/2$. Layer 2 has been implemented by a set of single-mode fiber beam splitters, whose reflectivities have been directly measured to be $R(\mathrm{BS}_{12}^{(2)}) \sim 0.515$, $R(\mathrm{BS}_{34}^{(2)}) \sim 0.507$ and $R(\mathrm{BS}_{56}^{(2)}) \sim 0.498$ respectively. Losses within the interferometer are mostly localized between layer 1 and 2, and are modeled as a fictitious BS of transmittivity $\eta_{i}$. Losses are due to fiber coupling from the source $(\eta \sim 0.4)$, fiber-to-fiber coupling in delay lines $(\eta \sim 0.7)$, and insertion losses in single-mode fiber beam splitters $(\eta \sim 0.7)$. Furthermore, additional losses $(\eta \sim 0.625)$ have been measured in mode 5 due to the presence of an extra fiber-to-fiber coupling stage.

Finally, the detection apparatus is composed of MMFBS and single-photon avalanche photodiodes (APD). The latter are non-photon number resolving detectors that click with probability $P_{\mathrm{det}}(\eta) = 1-(1-\eta_{\mathrm{det}})^{n}$ where $n$ is the number of impinging photons and $\eta_{\mathrm{det}}$ the detection efficiency ($\eta_{\mathrm{det}} \sim 0.6$). MMFBS are employed to obtain (probabilistically) photon-number resolution capabilities to the apparatus. The measured reflectivities for the employed beam splitter BS$_{i}^{(\mathrm{d})}$ were respectively (0.65, 0.77, 0.72, 0.77, 0.47, 0.15).

\section{Surface of the trivial lower bound for \MakeLowercase{$c_1$}}
In this section we discuss the model for the calculation of the surfaces, reported in the main text, corresponding to the conditions that lead to a trivial bound for $c_1$. In particular such calculation provides information on the temporal delays needed to find a non-zero contribution of genuine multiphoton interference. A simple way to deal with the multiphoton state for a given set of relative delays is to consider each single-photon state having the same spectrum, that is, a gaussian function centered at the same frequency $\omega_0$. Hence the state $\ket{\psi_i(t)}$ of a photon arriving at time $t$ in the interferometer can be expressed as follows:
\begin{equation}
    \ket{\psi_i(t)}=\int_{-\infty}^{+ \infty}d\omega \, \frac{1}{\sqrt{\pi}\Delta \omega} \, e^{-\frac{(\omega-\omega_{0})^2}{2 \Delta \omega^2}} \, e^{i \omega t} {a}^{\dagger}_{i,\omega}\ket{0}
\end{equation}
where $\Delta \omega$ is the spectrum's width and $a^{\dagger}_{i,\omega}$ is the bosons creation operator.
Then, the overlap between pairs of photons, in the same spatial mode $k$, is described by
\begin{equation}
\begin{aligned}
    r_{ij}(\Delta t_{ij})&=\left| \braket{\psi_k(t_i)}{\psi_k(t_j)}\right|^2=\\
&= e^{-\frac{1}{2}(t_i - t_j)^2\Delta \omega^2}=
    e^{-\frac{1}{2}\Delta t_{ij}^2\Delta \omega^2}
\end{aligned}
    \label{eq:overlaps}
\end{equation}

According to this model we can reconstruct the transition from the fully indistinguishable four-photon state to other configurations with different degree of distinguishability depending on the relative temporal delays. Consequently, we can also provide an expression for the lower bound of $c_1$, defined as the quantity $r_{AB}+r_{BC}+r_{CD}-2$ introduced in Eq.~(3) of the main text. To this aim, in order to furnish a proper description of the state for any relative delays, we can express it as a convex combination of all the possible configurations in which one and more photons in the four-tuple are fully distinguishable from the remaining others, according to the model (\textit{i}) described in the main text. For the present interferometer and input state, this complete ensemble includes only 8 configurations $s=\{$XXXX, XYYY, XYXX, XXYZ, XXXY, XXYY, XYYZ, XYWZ$\}$. Indeed the small number of different output distributions is due to the symmetry of the logical test that requires photons interfering pairwise. Then we have:
\begin{equation}
    \ket{\psi(\Delta t_{\mathrm{AB}}, \Delta t_{\mathrm{BC}}, \Delta t_{\mathrm{CD}})}=\mathcal{N}\sum_{s} \alpha^{\mathrm{AB}}_{s_{i}s_{j}}\alpha^{\mathrm{BC}}_{s_{j}s_{k}}\alpha^{\mathrm{CD}}_{s_{k}s_{l}}
    \,\rho_s,
    \label{eq:state}
\end{equation}
where $\mathcal{N}$ is a normalization constant, $\rho_s$ the density matrix associated to each configuration and the $\alpha$s are quantities related to the pairwise overlaps for a given delay. In particular, for instance considering the pair $($A, B$)$, they have the following expression:
\begin{equation}
\alpha^{\mathrm{AB}}_{s_{i}s_{j}}=
    \begin{cases}
 r_{\mathrm{AB}}(\Delta t_{\mathrm{AB}})& \text{if $s_{i}=s_{j}$} \\
1-r_{\mathrm{AB}}(\Delta t_{\mathrm{AB}})& \text{if $s_{i}\ne s_{j}$}
\end{cases}
\label{eq:alphas}
\end{equation}

The pink surface reported in Fig. 4 in the main text shows the resulting condition $c_1$ in the space of the relatives delays expressed as a function of $\Delta x_{ij}= c \Delta t_{ij}$. Such naive model for the four-photon state captures some features of the actual state produced by the source and by post-selection of the output states. Indeed the shapes of HOM dips in Fig. \ref{fig:dips} show that a gaussian spectrum is a good approximation to the actual one, which is in turn the result of the convolution between the spectra produced in the SPDC process and the narrow-band filters placed in the apparatus. During the calculation of the surface, we use as $\Delta \omega$ the widths of the measured HOM dips. However, the expression (\ref{eq:overlaps})  for the overlaps does not reflect the experimental dip visibilities $V_{ij}$, obtained during the preliminary characterization of the state in Section \ref{sec:part_ind}. Indeed the model predicts ideal overlaps for zero relative time delays. A step further in the model is to reformulate the expression (\ref{eq:overlaps}) so that $r_{ij}(0)=V_{ij}$. A possible solution is provided by the following expression
\begin{equation}
    r_{ij}(\Delta t_{ij})=V_{ij}
    e^{-\frac{1}{2}\Delta t_{ij}^2\Delta \omega^2},
    \label{eq:overlaps2}
\end{equation}
which, inserted in (\ref{eq:state}), produces the green surface in Fig. 4 of the main text. Note that this new surface explains why the configurations $\{$XXXY, XXYY$\}$ do not lie on the bound as predicted by the model for the pink surface. 
Furthermore, in the condition of zero delays, we obtain the following state:
\begin{equation}
\begin{aligned}
\rho_{source}&= 
V_{\mathrm{AB}}V_{\mathrm{BC}}V_{\mathrm{CD}}\rho_{\mathrm{XXXX}}+\\
&+(1-V_{\mathrm{AB}})V_{\mathrm{BC}}V_{\mathrm{CD}}\rho_{\mathrm{XYYY}}+\\
&+ V_{\mathrm{AB}}(1-V_{\mathrm{BC}})V_{\mathrm{CD}}\rho_{\mathrm{XXYY}}+\\
&+V_{\mathrm{AB}}V_{\mathrm{BC}}(1-V_{\mathrm{CD}})\rho_{\mathrm{XXXY}}+\\
&+V_{\mathrm{AB}}(1-V_{\mathrm{BC}})(1-V_{\mathrm{CD}})\rho_{\mathrm{XXYZ}}+\\
&+(1-V_{\mathrm{AB}})(1-V_{\mathrm{BC}})V_{\mathrm{CD}}\rho_{\mathrm{XYXX}}+\\
&+(1-V_{\mathrm{AB}})V_{\mathrm{BC}}(1-V_{\mathrm{CD}})\rho_{\mathrm{XYYZ}}+\\
&+(1-V_{\mathrm{AB}})(1-V_{\mathrm{BC}})(1-V_{\mathrm{CD}})\rho_{\mathrm{XYZW}}
\end{aligned}
\label{eq:state_exp}
\end{equation}
used as a model to obtain the theoretical output distributions for the ``XXXX" state.

The volume enclosed in each surface corresponds to the configurations where the input state is expected to possess a component of genuine indistinguishability. The ratio between the two volumes amounts to $\sim 0.525$. The actual ratio is even smaller if we take into account also the noise due to multi-pair emission in the SPDC, as explained in the previous section. However, it is worth noting that this limitation does not exist in the case of non-probabilistic single-photon sources such as quantum dots.

\section{Genuine four-photon indistinguishability and bounds on unmeasured overlaps}

 In this Section we show how to extend the results of \cite{Brod19,Brod19overlaps} to obtain the lower and upper bounds on the unmeasured overlaps, as well as on $c_1$ [i.e.\ Eqs. (3)-(10) in the main text]. In Sec.\ \ref{sec:classicalstates} we focus on the bounds obtained assuming a classical model for our photonic states, in which we have convex combinations of single-photon states which are pairwise identical or orthogonal (model ($i$) in the main text). In Sec.\ \ref{sec:nonclassicalstates} we obtain the bounds assuming a pure product-state model (model ($ii$) in the main text).

\subsection{Bounds for classical states} \label{sec:classicalstates}

\begin{figure}[h]
\centering
\includegraphics[width=0.4\textwidth]{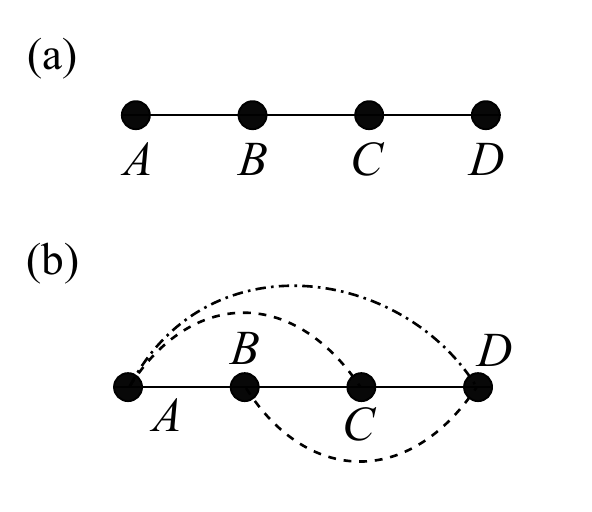}
\caption{{\bf Bounding unmeasured overlaps and $c_1$ from measurement overlaps}. (a) Graph encoding the known experimental data. Vertices are photons, and edges are measured two-photon overlaps. (b) By using the procedure we describe, we can ``complete'' the graph with ranges of possible values for the unmeasured overlaps.} 
\label{fig:4chain}
\end{figure}

Suppose we have a four-photon state given by a classical model, i.e.\ 
\begin{equation}
   \rho= c_1 \, \rho^{ind} + \sum_{s \neq 1} c_{s} \, \rho_{s}^{\perp} , \label{eq:class}
\end{equation}
where $\rho^{ind}$ is a state of four perfectly indistinguishable photons, and $\rho_{s}^{\perp}$ are pure states in a fixed basis with at least one pair of mutually orthogonal photons. We note that the state of Eq.\ (\ref{eq:state_exp}) is of this form. Suppose also that we have, as experimental data, the overlaps $r_{AB}, r_{BC},$ and $r_{CD}$ given by
\begin{equation}
    r_{ij} = 2 p_b^{ij} - 1
\end{equation}
where $p_b^{ij}$ is the probability of bunching in a HOM experiment between photons $i$ and $j$. 
We now show how to obtain bounds for the parameter $c_1$ in Eq.  (\ref{eq:class}) (identified in \cite{Brod19} as a measure of genuine four-photon indistinguishability) and the unmeasured overlaps $r_{AC}$, $r_{BD}$, and $r_{AD}$ (as discussed in \cite{Brod19overlaps}), in terms of the known two-photon overlaps. In the formalism of \cite{Brod19overlaps}, this situation corresponds to known two-state overlaps described by the $P_4$ graph shown in Fig.\ \ref{fig:4chain}(a).
Suppose initially that we have three logical propositions, $a_1$, $a_2$, and $a_3$, and let $p(a_i)$ be the probability that $a_i$ is true. Consider the truth table (\ref{table:and2}) for the AND function of the three propositions.
\begin{table}[h!]
  \begin{center}
    \begin{tabular}{|c|c|c|c|}
    \hline
      \textbf{$a_1$} & \textbf{$a_2$} & \textbf{$a_3$} & \textbf{$a_1 \wedge a_2 \wedge a_3$} \\
      \hline \hline
     0 & 0 & 0 & 0\\ \hline
     0 & 0 & 1 & 0\\ \hline
     0 & 1 & 0 & 0\\ \hline
     0 & 1 & 1 & 0\\ \hline
     1 & 0 & 0 & 0\\ \hline
     1 & 0 & 1 & 0\\ \hline
     1 & 1 & 0 & 0\\ \hline
     1 & 1 & 1 & 1\\ \hline
    \end{tabular}
\end{center}
\caption{Truth table for $a_1 \wedge a_2 \wedge a_3$.}
\label{table:and2}
\end{table}
We can now interpret each row in Table (\ref{table:and2}) as a four-dimensional vector. These vectors are the extremal points (vertices) of a polytope, and any logically consistent vector corresponds to a point in this polytope [i.e.\ a convex combination of the rows of Table (\ref{table:and2})]. The faces of this polytope correspond to linear logical inequalities that must be satisfied by any consistent set of probabilities \cite{Pitowsky94,Boole1854,Brod19overlaps}. It is a simple computational task to obtain such inequalities, which can be written as:
\begin{align}
p(a_1) &\le 1; \label{in1}\\
p(a_2) &\le 1; \label{in2}\\
p(a_3) &\le 1; \label{in3}\\
p(a_1 \wedge a_2 \wedge a_3 ) &\ge 0 ; \label{in4}\\ 
p(a_1 \wedge a_2 \wedge a_3 ) &\le p(a_1) ; \label{in5}\\ 
p(a_1 \wedge a_2 \wedge a_3 ) &\le p(a_2) ; \label{in6}\\ 
p(a_1 \wedge a_2 \wedge a_3 ) &\le p(a_3); \label{in7}\\
p(a_1 \wedge a_2 \wedge a_3 ) &\ge p(a_1)+p(a_2)+p(a_3)-2. \label{in8}
\end{align}
Inequalities (\ref{in1})--(\ref{in4}) are trivial. Inequalities (\ref{in5})--(\ref{in7}) encode the fact that the joint probability of several events must be at most the probability of any of the events individually. Inequality (\ref{in8}) is a nontrivial bound on $p(a_1 \wedge a_2 \wedge a_3 )$ in terms of the three individual probabilities.

Let us now interpret the three propositions as relations between pairs of photons connected by edges in Fig.\ \ref{fig:4chain}(a), i.e.,
\begin{align*}
a_1 := A = B, \\
a_2 := B = C, \\
a_3 := C = D. 
\end{align*}
Equalities in the above expressions mean that the two photons are in the same internal state (e.g.\ $A=B$ represents $r_{AB} = 1$). Since the state we consider at this point is of the form of Eq.\ (\ref{eq:class}), it is a classical mixture of states where each photon pair is sure to be either identical or orthogonal. Thus, we can directly interpret the probabilities of the propositions above as follows:
\begin{align}
r_{AB} &= p(a_1),\\
r_{BC} &= p(a_2), \\
r_{CD} &= p(a_3), \\
c_1 &= p(a_1 \wedge a_2 \wedge a_3).
\end{align}
Note that $c_1$ is interpreted as the probability that the four photons are identical, which matches its interpretation in terms of genuine indistinguishability given in \cite{Brod19}.
With this interpretation, inequalities (\ref{in5})--(\ref{in8}) lead immediately to 
\begin{align}
c_1 & \leq \min (r_{AB},r_{BC},r_{CD}),\\
c_1 & \geq r_{AB}+r_{BC}+r_{CD}-2,
\end{align}
which corresponds to inequality (3) in the main text.

We can also bound the values of the unmeasured overlaps. To that end, we can first repeat the argument above, but for the A-B-C subgraph of Fig.\ \ref{fig:4chain}(a). This leads straightforwardly to \cite{Brod19}
\begin{align}
r_{AB}+r_{BC}-1 \leq r_{AC} \leq  1-|r_{AB}-r_{BC}|,
\end{align}
which corresponds to inequality (4) from the main text. Equivalently, by looking at the B-C-D subgraph we obtain
\begin{align}
r_{BC}+r_{CD}-1 \leq r_{BD} \leq  1-|r_{BC}-r_{CD}|, \label{eq:boundBD}
\end{align}
which is inequality (5) from the main text.

From the bounds above we can effectively construct a new graph, where the ranges of allowed for values of $r_{AC}$ and $r_{BD}$ are added as new edges (i.e.\ the original graph together with the dashed edges in Fig.\ \ref{fig:4chain}(b)]. This new graph has a new 3-chain subgraph, namely A-B-D. If we apply the previous argument for this new subgraph, we obtain
\begin{align}
r_{AB}+r_{BD}-1 \leq r_{AD} \leq  1-|r_{AB}-r_{BD}|.
\end{align}
To further bound $r_{AD}$ in terms of known quantities, we just use the inequalities (\ref{eq:boundBD}) to extremize the bounds above over the full range of $r_{BD}$. After some algebra, this leads to
\begin{align}
r_{AD} \ge & r_{AB}+r_{BC}+r_{CD}-2,\\
r_{AD} \le & 2+ \min(r_{AB}-r_{BC}-r_{CD}, \\ &r_{AC}-r_{AB}-r_{CD}, r_{CD}-r_{AB}-r_{BC}),\notag
\end{align}
which correspond to inequalities (6) and (7) in the main text. Adding these new bounds to our graph leads to the complete graph of Fig.\ \ref{fig:4chain}(b) (including dashed and dot-dashed edges).

\subsection{Bounds for product pure states} \label{sec:nonclassicalstates}

We now obtain a new set of bounds when we model our experimental state as a pure product state. That is, suppose the state can be written as
\begin{equation}
    \ket{\psi} = \ket{A} \ket{B} \ket{C} \ket{D},
\end{equation}
how can we leverage the known overlaps $r_{ij} = \left \lvert \braket{i}{j}\right \lvert^2$ to infer something about the unmeasured overlaps?

For this, our starting point are the results of \cite{Brod19overlaps} for pure product states. There it was shown that, for a 3-state linear graph [e.g.\ for states A, B and C in the graph of Fig.\ \ref{fig:4chain}(a), though note we label the vertices differently than Ref.\ \cite{Brod19overlaps}], we have the following bounds
\begin{align}
r_{AC} &\le \left( \sqrt{r_{AB}r_{BC}}+\sqrt{(1-r_{AB})(1-r_{BC})}\right)^2, \label{eq:ub} \\
r_{AC} &\ge \left( \sqrt{r_{AB}r_{BC}}-\sqrt{(1-r_{AB})(1-r_{BC})}\right)^2 \label{eq:lb}.
\end{align}
The upper bound holds always. The lower bounds holds if the systems are qubits. If the systems are of dimension 3 or greater the lower bound only holds when $r_{AB} + r_{BC} >1$, otherwise the lower bound is trivial (i.e.\ 0). It was also shown in \cite{Brod19overlaps} that all of these bounds are achievable--that is, given fixed $r_{AB}$ and $r_{BC}$, and given any $r_{AC}$ in the range allowed by Eqs.\ (\ref{eq:lb}) and (\ref{eq:ub}), it is possible to find three pure states $\ket{A}$, $\ket{B}$ and $\ket{C}$ displaying that set of overlaps.

The above argument can be repeated to provide similar bounds for $r_{BD}$, namely,
\begin{align}
r_{BD} &\le \left(\sqrt{r_{BC}r_{CD}}+\sqrt{(1-r_{BC})(1-r_{CD})}\right)^2, \label{eq:ubBD} \\
r_{BD} &\ge \left( \sqrt{r_{BC}r_{CD}}-\sqrt{(1-r_{BC})(1-r_{CD})}\right)^2 \label{eq:lbBD}.
\end{align}
The bounds for $r_{AC}$ and $r_{BD}$ above correspond to Eqs.\ (8) and (9) from the main text.

Finally, we wish to leverage this construction to provide bounds for $r_{AD}$. By obtaining bounds for $r_{AC}$ and $r_{BD}$, we have effectively constructed a new graph, one represented by the solid and the dashed edges in Fig.\ \ref{fig:4chain}(b). Note that this new graph has a 3-vertex chain as a subgraph consisting of vertices ABD. This mean we can apply the same argument as above to bound $r_{AD}$ in terms of $r_{AB}$ and $r_{BD}$:
\begin{align}
r_{AD} &\le \left( \sqrt{r_{AB}r_{BD}}+\sqrt{(1-r_{AB})(1-r_{BD})}\right)^2, \label{eq:ubAD1} \\
r_{AD} &\ge \left( \sqrt{r_{AB}r_{BD}}-\sqrt{(1-r_{AB})(1-r_{BD})}\right)^2. \label{eq:lbAD1}
\end{align}
However, we already know the range of values that $r_{BD}$ can assume, from Eqs.\ (\ref{eq:lbBD}) and (\ref{eq:ubBD}). Therefore, the procedure to obtain the range of allowed values for $r_{AD}$ is to maximize Eq.\ (\ref{eq:ubAD1}) and minimize Eq.\ (\ref{eq:lbAD1}) over $r_{BD}$ in the corresponding interval. 

For the lower bound, an argument based on the monotonicity of the underlying two-variable function shows \cite{Brod19overlaps} that, if we take the lower bound of $r_{BD}$ from Eq.\ (\ref{eq:lbBD}), and substitute it into Eq.\ (\ref{eq:lbAD1}), this is an explicit (but cumbersome) expression for the lower bound of $r_{AD}$. However, for the corresponding upper bound we were not able to find an explicit expression. This is due to the fact, which we verified numerically, that the upper bound for $r_{AD}$ is obtained by sometimes using the lower bound and sometimes the upper bound for $r_{BD}$ into Eq.\ (\ref{eq:ubAD1}). 

It is also possible to bound $r_{AD}$ by considering vertices ACD rather than ABD, but a numerical investigation suggests that the resulting bounds are the same in both cases.